\documentclass[%
twocolumn,
superscriptaddress,
 amsmath,amssymb,
 aps,
pra
]{revtex4-1}

\usepackage{graphicx,subfigure}
\usepackage{color}
\usepackage{epstopdf}
\usepackage{dcolumn}
\usepackage{bm}
\usepackage{hyperref}

\begin{document}

\preprint{}

\title{Two-State Thermodynamics and the Possibility of a Liquid-Liquid
Phase Transition in Supercooled TIP4P/2005 Water}

\author{Rakesh S. Singh}
\affiliation{Department of Chemical \& Biological Engineering, Princeton University, Princeton, New Jersey 08544, USA}
\author{John W. Biddle}
\affiliation{Institute of Physical Science and Technology and Department of Chemical and Biomolecular Engineering, University of Maryland, College Park, Maryland 20742, USA}
\author{Pablo G. Debenedetti}
\affiliation{Department of Chemical \& Biological Engineering, Princeton University, Princeton, New Jersey 08544, USA}
\author{Mikhail A. Anisimov}
\email{anisimov@umd.edu}
\affiliation{Institute of Physical Science and Technology and Department of Chemical and Biomolecular Engineering, University of Maryland, College Park, Maryland 20742, USA}%

\date{\today}

\begin{abstract}
Water shows intriguing thermodynamic and dynamic anomalies in the supercooled liquid state. One possible explanation of the origin of these anomalies lies in the existence of a metastable liquid-liquid phase transition (LLPT) between two (high and low density) forms of water. While the anomalies are observed in experiments on bulk and confined water and by computer simulation studies of different water-like models, the existence of a LLPT in water is still debated. Unambiguous experimental proof of the existence of a LLPT in bulk supercooled water is hampered by fast ice nucleation which is a precursor of the hypothesized LLPT. Moreover, the hypothesized LLPT, being metastable, in principle cannot exist in the thermodynamic limit (infinite size, infinite time). Therefore, computer simulations of water models are crucial for exploring the possibility of the metastable LLPT and the nature of the anomalies. In this work, we present new simulation results in the $NVT$ ensemble for one of the most accurate classical molecular models of water, TIP4P/2005. To describe the computed properties and explore the possibility of a LLPT we have applied two-structure thermodynamics, viewing water as a non-ideal mixture of two interconvertible local structures (``states"). The results suggest the presence of a liquid-liquid critical point and are consistent with the existence a LLPT in this model for the simulated length and time scales. We have compared the behavior of TIP4P/2005 with other popular water-like models, namely mW and ST2, and with real water, all of which are well described by two-state thermodynamics. In view of the current debate involving different studies of TIP4P/2005, we discuss consequences of metastability and finite size in observing the liquid-liquid separation. We also address the relationship between the phenomenological order parameter of two-structure thermodynamics and the microscopic nature of the low-density structure.
\end{abstract}

\maketitle


\section{Introduction}
The peculiar behavior of supercooled water is still a puzzle that continues to attract strong interest. In contrast to the behavior of ordinary substances, the thermodynamic response functions (namely, the isothermal compressibility, the isobaric heat capacity and the magnitude of the thermal expansion coefficient) of liquid water show sharp increases on supercooling, suggesting their possible divergence just below the homogeneous ice nucleation temperature, $T_{\text{H}}$~\cite{angell_1973, speedy_1976, kanno_angell_1979, kanno_angell_1980, angell_1982, hare_1987, angell_sci, pablo_book, tombari_anomaly_1999, frank_book, stanley_pccp_2000, pablo_rev_2003}. Over the last few decades, several scenarios have been proposed to interpret this unusual thermodynamic behavior~\cite{taxeria_1980, speedy_1982, poole_1992, sf_1, sf_2, pablo_rev_2003, stanley_pnas_2005, bagchi_2011, tanaka_faraday}. One popular interpretation invokes a hypothetical first order liquid-liquid phase transition (LLPT) between two metastable liquid phases $-$ high-density liquid (HDL) and low-density liquid (LDL) $-$ a phenomenon known as \lq\lq liquid water polyamorphism\rq\rq~\cite{poole_1992}. The proposed liquid-liquid transition line terminates at a critical point in the supercooled region. Thus, the anomalous behavior of thermodynamic response functions in supercooled water is attributed to the hypothetical liquid-liquid critical point (LLCP) at which the response functions should diverge. Direct experimental observation of the LLPT and LLCP in bulk supercooled water is hampered by fast ice formation along $T_{\text{H}}$. Therefore, the hypothesized LLPT and LLCP are submerged in the \lq\lq no-man's land" (below $T_{\text{H}}$)~\cite{pablo_rev_2003}.

The known existence of two distinct glass transitions of water at ambient pressure is consistent with the hypothesized existence of two different forms of liquid water~\cite{pablo_rev_2003, mishima_nat_1998, giovam_2006, loerting_pnas_2013, giovam_2015}. Experiments on confined water (confinement is known to prevent crystallization) also suggest the existence of a LLPT in bulk water~\cite{mishima_nat_1998, mallamace_rev_2008}. However, due to strong effects of interaction of water with the surface of confining walls, the predictions of such studies may not be directly relevant to bulk water. Recently,  Sellberg \textit{et al.}~\cite{daniel_nat} used femtosecond X-ray laser pulses on micrometer-sized water droplets to probe the local structure of bulk supercooled water in ``no man's land" (water below $T_H$). Experiments on some supercooled aqueous solutions suggest the existence of a metastable liquid-liquid transition, presumably stemming from the original LLPT in pure supercooled water~\cite{Bruijn_2016,murata_2012, murata_2013, mishima_2014, zhao_angell_2016}, though the interpretation of experiments on glycerol-water solutions remains controversial~\cite{mishima_2014, biddle_2014, glycerol_water}.  In view of the enormous challenges that prevent direct access to the ``no-man's land", computer simulations remain crucial for exploring the structure, dynamics, and phase behavior of supercooled water.

Several computer simulation studies directly or indirectly suggest the existence of a metastable LLPT for some molecular models of water~\cite{poole_1992, st2_anomaly, pablo_2009, abascal_2010, pablo_2012, poole_2013, kesselring_2013, yagasaki_2014, pablo_nature_2014, anisimov_st2} and for tetrahedral network-forming models~\cite{sastry_nat_2003, franzese_2010, sastry_nat_2011, sciortino_natphys_2014, patchy_llcp}. Recent state-of-the-art free-energy computations convincingly confirm the LLPT for the ST2 model~\cite{poole_2013, pablo_nature_2014, small_st2} and some coarse-grained water-like network-forming models~\cite{sciortino_natphys_2014}.  Of special relevance is the work of Smallenburg and Sciortino~\cite{small_st2}: these authors showed that the LLPT in the ST2 model persists upon making the crystal phase metastable with respect to the liquid by tuning the hydrogen bond angular flexibility.  This disproves interpretations according to which the LLPT is generically a misinterpreted crystallization transition~\cite{limmer_1, limmer_2, limmer_c, kusalik_2013}.  On the other hand, metastable liquid-liquid separation is not observed in the coarse-grained mW model~\cite{mw,moore_2011,limmer_1,anisimov_mb}, a re-parametrized version of the Stillinger-Weber (SW) model~\cite{sw}.  Placing phenomenological (equation of state) calculations on firmer theoretical~\cite{anisimov_st2} and computational ground~\cite{pablo_nature_2014}, understanding the molecular basis underlying the existence~\cite{pablo_nature_2014} or absence~\cite{moore_2011} of a LLPT in specific models, and elucidating the model-dependent time and length scales over which a metastable LLPT can be observed~\cite{patey_2012, patey_2013, patey_2015} are current objects of activity and robust discussion.

The main focus of our work is the TIP4P/2005 water model~\cite{tip4p}. This model reproduces satisfactorily the thermodynamics of liquid water and the complex, experimentally observed phase diagram of water in its numerous crystalline phases, and is considered to be one of the most accurate classical molecular models of liquid water. The existence of a LLPT in the TIP4P/2005 model is still a subject of debate. Abascal and Vega~\cite{abascal_2010} reported the existence of a LLPT with critical parameters $T = 193$~K, $\rho = 1012$~kg/m$^3$ and $P = 135$~MPa from molecular dynamics (MD) simulations in the $NPT$ ensemble.  The more recent study of Sumi and Sekino~\cite{sumi_2013} in the $NPT$ ensemble also suggests a LLPT for the TIP4P/2005 model, but the critical parameters ($T \approx 182$ K, $\rho \approx 1020$ kg/m$^3$ and $P = 1580-1620$ bar) were found to be significantly different than those proposed by Abascal and Vega~\cite{abascal_2010}. An equation of state based on the concept of the presence of two different local structures, proposed by Russo and Tanaka~\cite{tanaka_natcomm}, also suggests the existence of a metastable LLPT for this model.  Recently, Yagasaki \textit{et al.}~\cite{yagasaki_2014} have carried out MD simulations in the $NVT$ ensemble and have observed a spontaneous low- and high-density liquid-liquid phase separation in three models: ST2, TIP5P, and TIP4P/2005. The critical temperature and density of TIP4P/2005 reported by Yagasaki \textit{et al.}~\cite{yagasaki_2014} are in agreement with the predictions of Sumi and Sekino~\cite{sumi_2013}. These authors also observed a clear separation of time scales between LLPT and crystallization. However, studies of Limmer and Chandler~\cite{limmer_2, limmer_c} and Overduin and Patey~\cite{patey_2013, patey_2015} found no evidence for two metastable liquid phases around the temperature-pressure range suggested by Abascal and Vega~\cite{abascal_2010}. Overduin and Patey~\cite{patey_2013} also argued that the simulations of Abascal and Vega~\cite{abascal_2010} are too short to obtain converged results. In recent studies, Limmer and Chandler~\cite{limmer_c} and Overduin and Patey~\cite{patey_2015} have also challenged the results of Yagasaki \textit{et al.}

Specifically, for TIP4P/2005~\cite{tip4p} and TIP5P~\cite{tip5p}, Overduin and Patey~\cite{patey_2015} show that the spontaneous liquid-liquid phase separation reported by Yagasaki \textit{et al.}~\cite{yagasaki_2014} exhibits a strong system size dependence. For a system size of $4000$ molecules, both studies, Ref.~\cite{yagasaki_2014} and Ref.~\cite{patey_2015}, observe regions of different densities separated by well-defined planar interfaces. However, Overduin and Patey~\cite{patey_2015} also observed that the density difference between these regions was sharply reduced with increasing system size, and disappeared for a system size of $32000$ molecules. These authors further argue that, as the appearance of regions of low density is always accompanied by an excess of local ice-like molecules, the regions of different densities observed by Yagasaki \textit{et al.}~\cite{yagasaki_2014} are likely associated with appearance and coarsening of local ice-like structures, rather than with liquid-liquid phase separation. This argument supports the conclusion of Limmer and Chandler~\cite{limmer_c}, who also argued that the density differences observed by Yagasaki \textit{et al.}~\cite{yagasaki_2014} are due to ice coarsening. 

The fact that metastable phase behavior depends on the system size is not surprising. Obviously, a metastable phase separation cannot exist in the thermodynamic limit (infinite size and infinite time of equilibration). This is why the results and arguments of Overduin and Patey~\cite{patey_2015}, as well as of Limmer and Chandler~\cite{limmer_1, limmer_2, limmer_c}, require thorough analysis in light of the physics of metastability.

There is another aspect of the physics of supercooled water that is closely related to the discussion of the possibility of a metastable LLPT. This has to do with the physical nature of the thermodynamic anomalies, in particular, the trend toward diverging response functions. Currently, there is broad consensus based on the experimental~\cite{soper_2000, torre_natcomm, lars_2008, lars_2009, lars_2015, Nilsson_2015} and simulation~\cite{lars_2011, lars_rev, moore_2009, poole_d5, tanaka_natcomm} studies, that in supercooled water two competing local structures indeed exist. Could this competition, which is assumed to be responsible for the thermodynamic anomalies, be sharp enough to trigger a metastable LLPT? This is the central question. Definitely, this possibility is strongly model-dependent and could also depend on specific experimental/simulation conditions.

Recently, two-structure thermodynamics has become increasingly popular for explaining the anomalous thermodynamic behavior of supercooled water~\cite{pablo_nature_2014, anisimov_nat, anisimov_mb, anisimov_st2, bertrand, lars_2015, tanaka_2000, tanaka_natcomm}. Liquid water is considered as a \lq\lq mixture" of two types of local environments $-$ LDL-like and HDL-like, with the fraction of each controlled by thermodynamic equilibrium. The competition between these two distinct configurations naturally explains the density anomaly along with other thermodynamic anomalies in the supercooled state. If the excess Gibbs energy of mixing of these two structures is positive, the non-ideality of mixing can overcome the ideal entropy of mixing, causing liquid-liquid phase separation. Recent studies show that the thermodynamic properties of metastable liquid water~\cite{anisimov_nat}, as well as of the ST2~\cite{anisimov_st2} and mW~\cite{anisimov_mb} water models, can be well described by two-structure thermodynamics. It was shown that the liquid-liquid phase separation for the ST2 model is energy-driven, however, for the mW model, non-ideality of mixing is only entropy driven and is not strong enough to induce a LLPT. Bresme \emph{et\hspace{1mm}al.} used a two-structure equation of state with a LLPT and LLCP to describe the TIP4P/2005 model in an investigation of the model's thermal conductivity, and found good agreement between the model and the simulation data \cite{Bresme_2014}. So far, the best description of all currently available experimental data on thermodynamic properties of supercooled water is achieved by an equation of state based on two-structure thermodynamics~\cite{anisimov_nat, anisimov_new}. A semi-empirical extension (up to 400~MPa) of this equation of state, reported in Ref.~\cite{anisimov_new}, has been adopted by the International Association for the Properties of Water and Steam (IAPWS) as an international guideline for scientific and industrial use. The recently observed bimodal distributions of molecular arrangements of inherent structures in the SPC/E~\cite{lsi_spce} and TIP4P/2005~\cite{lars_2011} models strongly support the two-structure description of liquid water. The existence of a bimodal distribution of molecular configurations in water is also supported by X-ray absorption and emission spectroscopy~\cite{lars_2015, lars_2008} and by an investigation of vibrational dynamics~\cite{torre_natcomm}. However, the mere existence and competition of the two local structures in water do not necessarily mean the existence of a metastable LLPT in \lq\lq no-man's land"~\cite{tanaka_faraday}.

In this work, we have carried out extensive computer simulations in order to explore the nature of the thermodynamic anomalies and, consequently, the possibility of a metastable LLPT in TIP4P/2005. To describe the computed properties, we have applied two-structure thermodynamics, viewing water as a non-ideal mixture of two interconvertible local structures. The thermodynamic behavior of the model in the one-phase region is fully consistent with the existence of an energy-driven LLPT in this model (at least for the simulated length and time scales). We have compared the behavior of TIP4P/2005~\cite{tip4p} with the mW~\cite{mw} and ST2~\cite{st2} models, and with real water. We have also addressed the relation between the phenomenological order parameter of two-state thermodynamics and the microscopic nature of the low-density structure. In view of the current controversy between different studies of TIP4P/2005, the crucial role of metastability and finite size in observing liquid-liquid separation is emphasized.

\section{Computational Model and Simulation Details}
We performed molecular dynamics (MD) simulations of $216$ water molecules interacting via the TIP4P/2005 pair potential~\cite{tip4p} in a cubic box at constant temperature and volume ($NVT$ ensemble). We computed the properties of liquid water at approximately $200$ state points at densities ranging from $1120-960$ kg/m$^3$ in steps of $20$ kg/m$^3$ and temperatures ranging from $300$ K down to $185-180$ K (depending on the density of the system) in steps of $5$ K. This choice of ensemble was partly motivated by the possibility of observing van der Waals loops in the two-phase region (below LLCP), in case they exist. It turned out that we were not able to relax the system in the region where one would expect to observe van der Waals loops. However, using many state points in the $NVT$ ensemble enables us to follow different isochores throughout the one-phase region, and to extrapolate them into the region where the slow relaxation of the system impedes reliable computation.  We may thus distinguish between a system with a LLPT, in which the isochores are projected to cross, and a system with competition between two structures but without a LLPT, in which the isochores do not cross.

We have also performed MD simulations in the $NPT$ ensemble with $512$ water molecules at 0.1~MPa and temperatures ranging from $300$ K to $200$ K. All simulations were performed with use of GROMACS $4.6.5$ molecular dynamics simulation package~\cite{gromacs}. In all cases, periodic boundary conditions were applied, and a time step of $2$ fs was used. The short-range interactions were truncated at $8.5$~\AA~ for $216$ water molecule system and $9.5$~\AA~ for $512$ water molecule system. Long range electrostatic terms were computed by particle mesh Ewald with a grid spacing $1.2$~\AA. Long range corrections were applied to the short range Lennard-Jones interaction for both energy and pressure. Bond constraints were maintained using the
LINCS algorithm~\cite{lincs}. To maintain constant temperature we used a Nose-Hoover thermostat~\cite{nose, hoover} with $0.2$ ps relaxation time. Constant pressure was maintained by a Parrinello-Rahman barostat~\cite{pr_barostat} with $2$ ps relaxation time. 

Molecular models of water are notorious for extremely slow structural relaxation in the superooled state. This slow structural relaxation often leads to controversy over thermodynamic behavior of supercooled water observed in computer simulation studies~\cite{pablo_2012, limmer_1, limmer_2, patey_2013}. In this work, in order to ensure the relaxation of the system at each state point, we computed and carefully monitored the decay of the self part of the intermediate scattering function ($F_s(k,t)$)~\cite{hansen_book} with time $t$ (shown and discussed in the Appendix). To ensure the relaxation of the system at each thermodynamic condition investigated in this work, MD trajectories are at least 400 times as long as the structural relaxation time (defined as the time at which $F_s(k^*,t)$ = 1/e, $k^*$ is the wavenumber corresponding to the first peak of structure factor). Depending on the thermodynamic condition, MD trajectory lengths vary between 20 ns and 15 $\mu$s.

\section{Thermodynamics of two states in liquid water}\label{tseos}
The two-structure equation of state (TSEOS) treats liquid water as a \lq\lq mixture" of two interconvertible structures (``states"), a high-density/high-entropy structure $A$ and a low-density/low-entropy structure $B$. These two structures are interconvertible by a process that can be viewed as analogous to a ``chemical reaction" $A \rightleftarrows B$. This means that, unlike in binary mixtures, the fraction of each structure, $1-x$ for $A$ or $x$ for $B$, is not an independent variable but rather is controlled by thermodynamic equilibrium. Our expression for the molar Gibbs energy of the system takes the form~\cite{anisimov_st2, anisimov_mb, anisimov_nat, anisimov_new}:
\begin{equation}
G = G^A + x G^{BA} + RT \left[ x \ln x + (1-x) \ln (1-x) + \omega x (1-x) \right],
\end{equation}
where $G^A$ represents the Gibbs energy of pure structure $A$ and $G^{BA} = G^B - G^A$ represents the difference in Gibbs energy between structures $B$ and $A$, respectively. The term $G^A$ is treated empirically as a polynomial function of temperature and pressure. Since we are testing the possibility of the existence of a LLPT (terminating at a LLCP) for the TIP4P/2005 model, the convenient variables are $\Delta\widehat{T}=(T-T_c)/T_c$ and $\Delta\widehat{P}=(P-P_c)/\rho_cRT_c$, where $T$ is the temperature, $P$ is the pressure, $T_c$, $P_c$ and $\rho_c$ are the critical temperature, pressure, and molar density, respectively. Therefore, $G^A$ is represented as
\begin{equation}
G^A = \sum_{m,n}c_{mn} \Delta\widehat{T}^m \Delta\widehat{P}^n,
\end{equation}
with $\{c_{mn}\}$ being adjustable coefficients. The difference $G^{BA}$ determines the equilibrium constant $K$ of the ``chemical reaction" $A \rightleftarrows B$ as $\ln(K(T,P))=-G_{BA}/{RT}$. In the simplest non-linear approximation, 
\begin{equation} \label{eqn:3}
\frac{G^{BA}}{RT} = \lambda{(\Delta\widehat{T}+a\Delta\widehat{P} + b\Delta\widehat{T}\Delta\widehat{P})},
\end{equation}
where $\lambda$ is associated with the difference in entropy between the two structures, $a$ gives the slope $-(dT/dP)$ of the LLPT at the critical point and thus, asymptotically, the slope of the critical isochore, and $b$ gives the curvature of the LLPT line and its analytic continuation, the \lq\lq Widom line"\cite{anisimov_st2,anisimov_nat}. The condition $\ln K = 0$ describes the LLPT, LLCP, and the Widom line.
  
The fraction $x$ of molecules associated with structure $B$ is controlled by the condition that the value of the Gibbs energy must be a minimum in thermodynamic equilibrium, so the equilibrium fraction $x_e$ can be found from the equation
\begin{equation}  \label{eqn:xe}
\left(\frac{\partial G(T,P;x)}{\partial x}\right)_{T,P} = 0.
\end{equation}

In an ideal mixture, $x_e$ will vary smoothly with $T$ and $P$ and there will be no phase transition~\cite{tanaka_faraday}.  If, however, the mixture is sufficiently non-ideal, the change in $x_e$ may be discontinuous, signifying a first order phase transition between a high-density liquid (rich in structure $A$) and a low-density liquid (rich in structure $B$).  In our case, the variation of the non-ideality yields a phase diagram with a curve of first-order LLPT terminating at a LLCP. An extension of the LLPT curve into the one-phase region, where the non-ideality is not strong enough to induce phase separation, is commonly called the Widom line~\cite{stanley_pnas_2005}. Asymptotically close to the critical point, the Widom line coincides with the critical isochore, as well as with loci of maxima in the isobaric heat capacity $C_P$ and isothermal compressibility $\kappa_T$~\cite{stanley_pnas_2005}.

The term $RT \left[ x\ln x + (1-x)\ln (1-x) + \omega x (1-x) \right]$ describes the Gibbs energy of mixing of structures $A$ and $B$ with $x\ln x + (1-x)\ln(1-x)$ being the ideal entropy of mixing~\cite{swinton_rowlinson}. The non-ideality is represented by a simple form, symmetric in $x$, with the parameter $\omega$ determining the nature and strength of the non-ideality.  In this simple (symmetric) form of the TSEOS, the critical composition $x_c = 1/2$. If $\omega$ does not depend on temperature, then the non-ideality that leads to phase separation is entirely due to the non-ideal entropy of mixing (\lq\lq athermal solution"). This form of the non-ideality has been used to describe both real-water~\cite{anisimov_nat} and mW water~\cite{anisimov_mb}. If, on the other hand, $\omega \propto 1/T$, then the non-ideality arises due to non-ideal enthalpy of mixing (\lq\lq regular solution").  Such a ``regular solution" model has been used to describe two versions of the ST2 model of water, ST2(I) and ST2(II)~\cite{anisimov_st2}.  

In principle, both non-ideal entropy and enthalpy of mixing could contribute. However, for TIP4P/2005, modification of the temperature dependence of $\omega$, which is equivalent to the inclusion of non-ideal entropy of mixing, introduced an additional adjustable parameter but did not yield an improvement in the description of the simulation data within their uncertainties.  Therefore, for simplicity, we model the non-ideality as arising only from non-ideal enthalpy of mixing, as was done for ST2~\cite{anisimov_st2}. In this model, phase separation occurs for $\omega > 2$, so we give $\omega$ the linear form
\begin{equation} \label{eq5}
\omega = \frac{2 + \omega_0\Delta \hat{P}}{\hat{T}},
\end{equation}
where $\hat{T} = T/T_c$, and $\omega_0$ is the only adjustable parameter that controls the non-ideality of mixing.

Simple approximations, given by Eq.~\ref{eqn:3} and Eq.~\ref{eq5}, used for describing the TIP4P/2005 model, while enabling us to avoid a large number of adjustable parameters, obviously restrict the validity of the TSEOS within a reasonable vicinity of the LLCP. In this work, we are deliberately using this restriction to describe the area of converging isochores and thus to emphasize the possibility of the existence of LLPT in this model.

To obtain the values of thermodynamic properties from the TSEOS we first solve Eq.~\ref{eqn:xe} for $x_e$, and then use $x_e$ to evaluate the desired derivative of the Gibbs energy. Because Eq.~\ref{eqn:xe} is a transcendental equation with no closed-form solution, numerical methods must be used.
\section{Description of thermodynamic properties of TIP4P/2005 water}
\begin{figure}[htbp!] 
\vspace{10pt}
\includegraphics[width=0.45\textwidth]{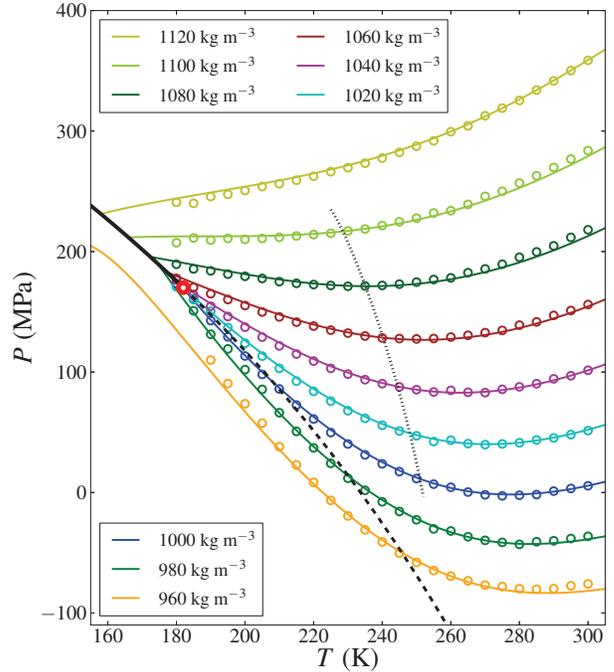}
\caption{Isochores in the $P-T$ plane for TIP4P/2005 model. The open circles indicate simulation data, while the solid lines show the same iscohores according to the TSEOS. The LLPT line, the LLCP, and the Widom line are shown as the solid black line, large red circle, and black dashed line, respectively.  The thin dotted line is the melting line of TIP4P/2005 as reported in Ref.~\cite{abascal_2010}.}
\label{fig1}
\end{figure}

\begin{figure}[htbp!] 
\vspace{10pt}
\includegraphics[width=0.47\textwidth]{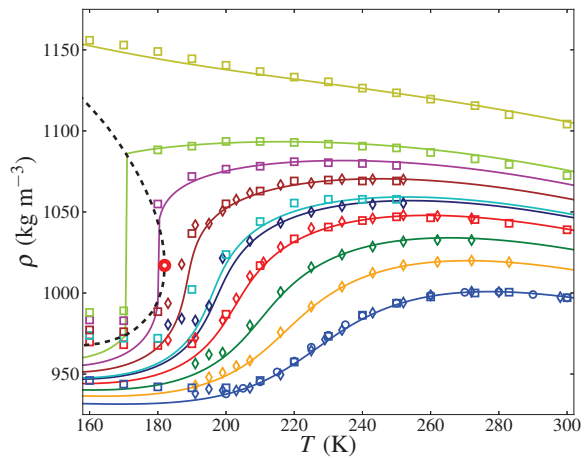}
\caption{Densities along isobars computed by Sumi and Sekino~\cite{sumi_2013} (open squares), Abascal and Vega~\cite{abascal_2010} (open diamonds), in this work (open circles; 0.1~MPa), and fits by the TSEOS (solid lines).  The black dashed line bounds the two phase region as predicted by the TSEOS, and the red circle shows the predicted location of the critical point.  Isobars shown, from top to bottom, are 300, 200, 175, 150, 125, 120, 100, 70, 40, and 0.1~MPa.}
\label{fig2}
\end{figure}

In Fig.~\ref{fig1}, we present the results of our $NVT$ simulations (open circles) along with isochores predicted by the TSEOS (solid lines). The densities range from $960$ to $1120$ kg/m$^3$ in steps of $20$ kg/m$^3$ and the temperatures range from $300$ K down to $180-190$ K (depending on the density) in steps of $5$ K. The error bars of the simulation data points are approximately equal to the size of the circles. The shape of the isochores in the supercooled region strongly suggests the existence of a LLCP for this model as predicted by the TSEOS. In Fig.~\ref{fig2}, we have  compared TSEOS predictions for the densities along isobars with the previously reported data by Sumi and Sekino~\cite{sumi_2013} as well as by Abascal and Vega~\cite{abascal_2010} obtained by $NPT$ simulations. We observe quantitatively good agreement between predictions of the TSEOS and simulation data for $T > 200$ K. However, for very low temperatures, the densities predicted by the TSEOS deviate significantly from previously reported data~\cite{sumi_2013}. This discrepancy most likely arises due to the approximations used in the current form of the TSEOS.  In any case, both the new simulation data and the TSEOS strongly imply the existence of a liquid-liquid critical point near 182~K and 170~MPa, consistent with the recent simulation studies by Yagasaki \textit{et al.}~\cite{yagasaki_2014} and by Sumi and Sekino~\cite{sumi_2013}.

In Fig.~\ref{fig1}, and Fig.~\ref{fig2} we show the TSEOS prediction for the LLPT and LLCP in the $P-T$ and $\rho-T$ planes, respectively. The TSEOS parameters are reported in Table~\ref{fit_data}. The proposed phase diagram and its comparison with other studies for this model are discussed in more detail in Section~\ref{phase}.
\begin{table}[ht]
\caption{Parameters for the two-structure equation of state \footnote{The adjustable coefficients of Eq. (2) are made dimensionless by the critical parameters}}
\centering 
\begin{tabular}{l@{\hskip 0.3in}c@{\hskip 0.3in}c@{\hskip 0.3in}c c}
\hline
\hline                        
Parameter & Value & Parameter & Value \\ [1ex]
\hline
$T_c$ & $182$ K & $\hat{c}_{20}$ & $-5.3481$  \\ 
$P_c$ & $170$ MPa & $\hat{c}_{12}$ & $0.000493$  \\
$\rho_c$ & $1017$ kg/m$^3$ & $\hat{c}_{21}$  & $0.1094$  \\
$\lambda$ & $1.407$ & $\hat{c}_{30}$ & $1.3293$ \\
$a$ & $0.171$ & $\hat{c}_{22}$ & $-0.02129$  \\ 
$b$ & $-0.100$ & $\hat{c}_{31}$ & $-0.02446$ \\ 
$\omega_0$ & $0.0717$ & $\hat{c}_{40}$ & $-0.13173$ \\ 
$\hat{c}_{01}$ & $0.8617$ & $\hat{c}_{23}$ & $0.003687$ \\ 
$\hat{c}_{02}$ & $-0.003412$ & $\hat{c}_{32}$ & $0.01229$  \\ 
$\hat{c}_{11}$ & $0.01351$ & $\hat{c}_{33}$ & $-0.003513$ \\ [1ex]
\hline
\hline
\end{tabular}
\label{fit_data}
\end{table}

\begin{figure}[htbp!]
 \vspace{15pt}
        \subfigure{
            \label{fig3a}
            \includegraphics[width=0.48\textwidth]{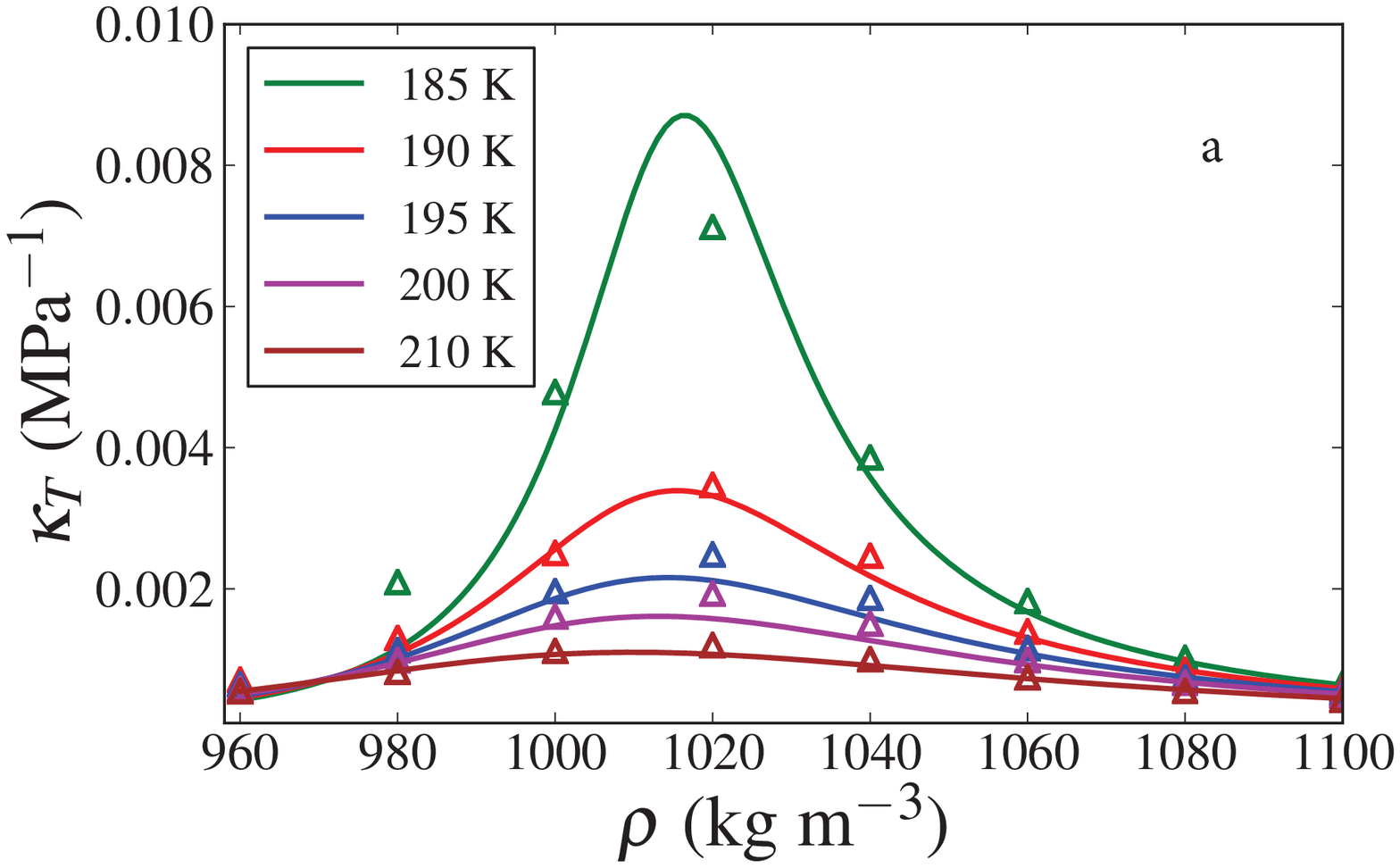}
        }
        \subfigure{
           \label{fig3b}
           \includegraphics[width=0.475\textwidth]{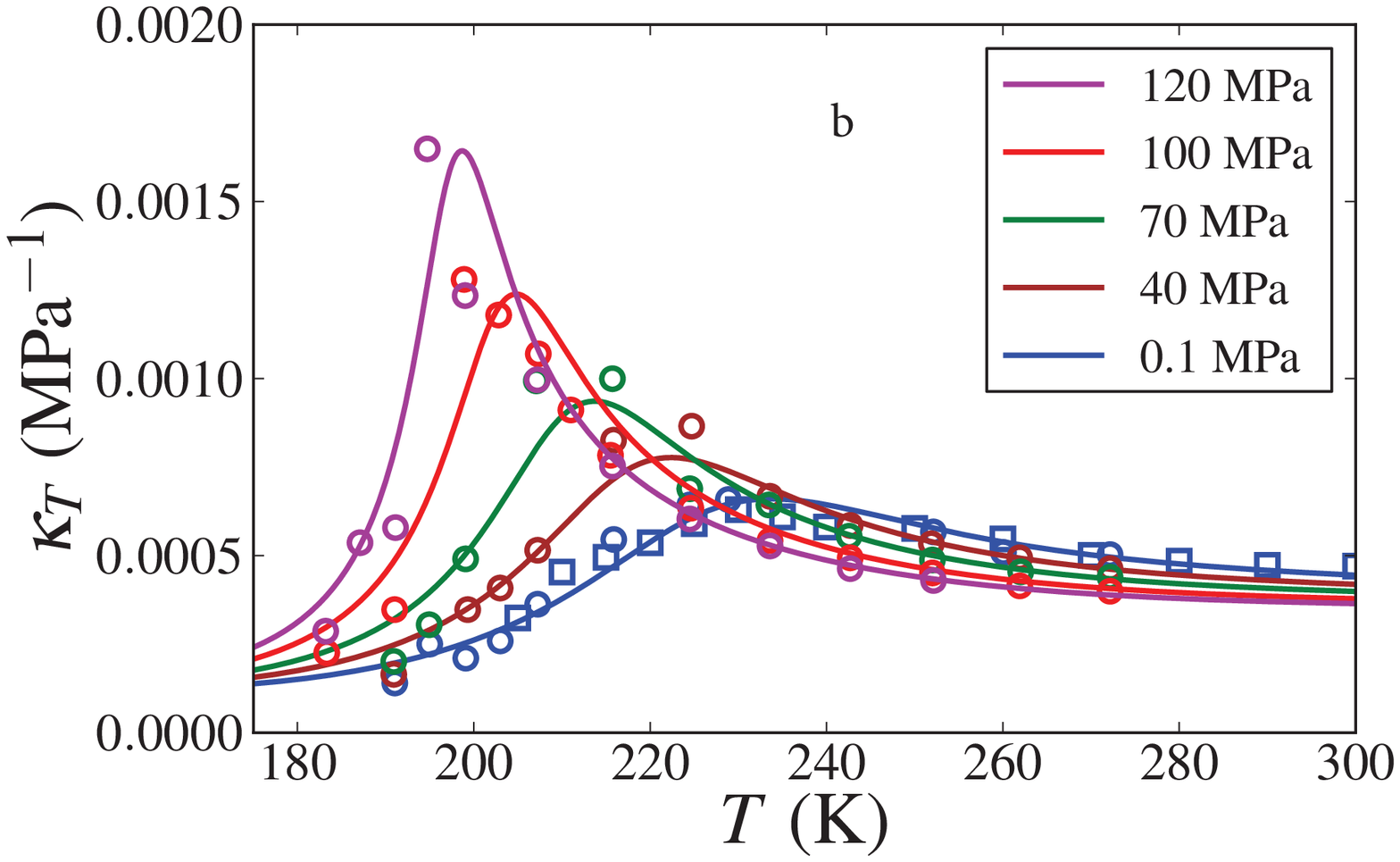}
        }
    \caption{(a) Isothermal compressibility along isotherms. Symbols are simulation data and the curves are predictions of the TSEOS (this work). (b) Isothermal compressibility along isobars. Symbols are simulation data by Abascal and Vega~\cite{abascal_2010} (open circles) along with our work at $0.1$ MPa (open squares). The curves are the predictions by the TSEOS.}
   \label{fig3}
\end{figure}

In order to gain deeper insight into the thermodynamic behavior of the TIP4P/2005
model in the supercooled state, in Fig.~\ref{fig3a} and Fig.~\ref{fig3b} we demonstrate the behavior of the isothermal compressibility($\kappa_T$) and the corresponding predictions of the TSEOS. The isothermal compressibility, defined as, $\kappa_{T} = -(1/V)(\partial{V}/\partial{P})_{T} = \langle(\delta{V})^2\rangle / k_{\text{B}}TV$ ($k_{\text{B}}$ is Boltzmann's constant, $V$ is the volume), is a measure of the mean-square volume fluctuations $ \langle(\delta{V})^2\rangle$ at constant temperature. The compressibility as a function of density along isotherms, shown in Fig.~\ref{fig3a}, is computed from the simulation data. In Fig.~\ref{fig3b}) we show the compressibility data along the isobars reported by Abascal and Vega~\cite{abascal_2010} to compare with the predictions of the TSEOS. The results presented in these figures show that the two-structure thermodynamics successfully describes the observed anomalous behavior of thermodynamic response functions in the supercooled region.     

\begin{figure}[htbp!]
 \vspace{10pt}
\includegraphics[width=0.45\textwidth]{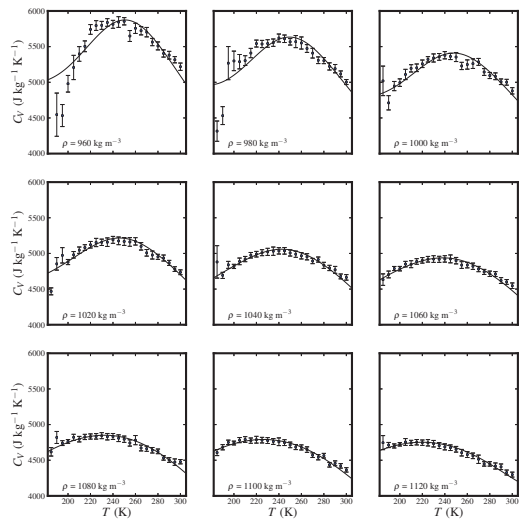}
    \caption{Temperature dependence of the specific heat capacity at constant volume ($C_V$) along different isochores. Symbols are simlutation data computed from total energy fluctuations and solid lines show the predictions of the TSEOS.}
   \label{fig4}
\end{figure}

In Fig.~\ref{fig4}, we demonstrate the temperature dependence of the heat capacity at constant volume, $C_V$, along different isochores. The isochoric heat capacity $C_V$, defined as $(\partial{E}/\partial{T})_V = \langle(\delta{E})^2\rangle / k_{\text{B}}T^2$, is a measure of the total energy ($E$) fluctuations of the system. We observe excellent agreement between two-state thermodynamics and computer simulation predictions for higher densities (greater than $1040$ kg/m$^3$) and reasonably good agreement (considering the larger uncertainties involved in computing energy fluctuations) at lower densities and temperatures. From the figure, it is evident that unlike $\kappa_T$, $C_V$ does not show any significant anomaly on supercooling down to $185$ K. This is not surprising, as the anomaly of the isochoric heat capacity near the critical point is very weak.  It originates solely from fluctuation effects associated with the divergence of the correlation length and does not exist in the mean-field approximation. According to scaling theory~\cite{fisher_1983}, the weak divergence of $C_V$ should only be noticeable in the close vicinity of the critical point (practically, within $1$-$2$ degrees, \emph{i.\hspace{1mm}e.} at $(T-T_c)/T_c < 10^{-2}$~\cite{anisimov_book}).  Moreover, in a finite-size system, the correlation length cannot exceed the size of the box, and so the critical anomalies are rounded. Our system contains only $216$ molecules, which is too small for weak (fluctuation-induced) anomalies to be observed.  A crossover TSEOS that incorporates fluctuation effects upon approaching the critical point has recently been applied for the description of the ST2 model~\cite{anisimov_st2}, and it was shown that within the accuracy of simulation data for that model, fluctuation effects are negligible.

\section{Low-density fraction and the nature of the order parameter}
Two-structure thermodynamics makes use of the fraction $x$, an extent of reaction between two interconvertible structures, as a phenomenological order parameter.  It does not, however, specify the microscopic nature of these two structures, nor does it give a microscopic definition of the order parameter.  Different authors have suggested various ways to discriminate between the two arrangements of molecules in water~ \cite{tanaka_2000,pablo_nature_2001,mw,tanaka_natcomm,st2_anomaly,poole_d5}.

In this work, we have computed the order parameter based on two different criteria: distance to the fifth nearest neighbor ($d_5$)~\cite{poole_d5} and local structure index (LSI, usually denoted by $I$)~\cite{lsi}.  

The $d_5$ criterion assigns molecules to belong to \lq\lq low density" when $d_5$ is greater than the cut-off distance, $r_0 = 3.5$~\AA. This cut-off distance defines the first coordination shell and is estimated from the position of the minimum that separates the first and second coordination shells in the oxygen-oxygen radial distribution function. The parameter $d_5$ contains information about the local structure up to first coordination shell ($3.5$~\AA) only. 

In order to include structural information beyond the first shell, we have also computed the local structural index (LSI). The LSI of molecule $i$ is obtained by ordering the oxygen-oxygen nearest neighbor distances between the central $i^{th}$ molecule and its $j^{th}$ nearest neighbor (denoted as $r_j$); $r_1 < r_2 < ... ~r_j~ ....< r_n(i)<3.7$~\AA$ < r_{n(i)+1}$. The number $n(i)$ is chosen in such a way that
$r_{n(i)} < 3.7$~\AA$ < r_{n(i)+1}$. Then, LSI is defined as~\cite{lsi}
\begin{equation}
 I(i) = \frac{1}{n(i)}\sum_{j=1}^{n(i)}\left[\Delta(j;i) - \bar\Delta(i)\right]^2,
\end{equation}
where $\Delta(j;i) = r_{j+1} - r_j$ and $\bar\Delta$ is the average of $\Delta(j;i)$ over all the nearest neighbors $j$ of molecule $i$. LSI is a measure of inhomogeneity between the first and second hydration shells of a tagged water molecule and thus probes the local structure beyond the first shell. A large value of LSI implies that there is a structured first shell and that there is no inhomogeneity (that is, no trapped \lq\lq interstitial\rq\rq water molecules) between the first and second coordination shells. A small value of LSI implies either a disordered first coordination shell or a significant presence of inhomogeneities in between first and second coordination shells.  We have used the same procedure followed by Wikfeldt \textit{et al.}~\cite{lars_2011} for TIP4P/2005 and Appignanesi \textit{et al.}~\cite{lsi_spce} for SPC/E water to define LDL and HDL-like local environments in the system.  However, unlike these studies, we have computed the LDL fraction in real dynamical trajectories, not in the inherent structures. The particles having LSI value less than $0.13$~\AA$^2$ are assigned as HDL-like and particles having LSI values greater than $0.13$~\AA$^2$ as LDL-like.  The parameter $d_5$ requires merely that the low-density structures have a four-coordinated first shell, while the LSI criterion also requires local ordering beyond first coordination shell.  Consequently, low-density fraction as computed by the LSI criterion will in general be lower than that computed according to the $d_5$ criterion.

\begin{figure}[htbp!] 
\vspace{10pt}
\includegraphics[width=0.47\textwidth]{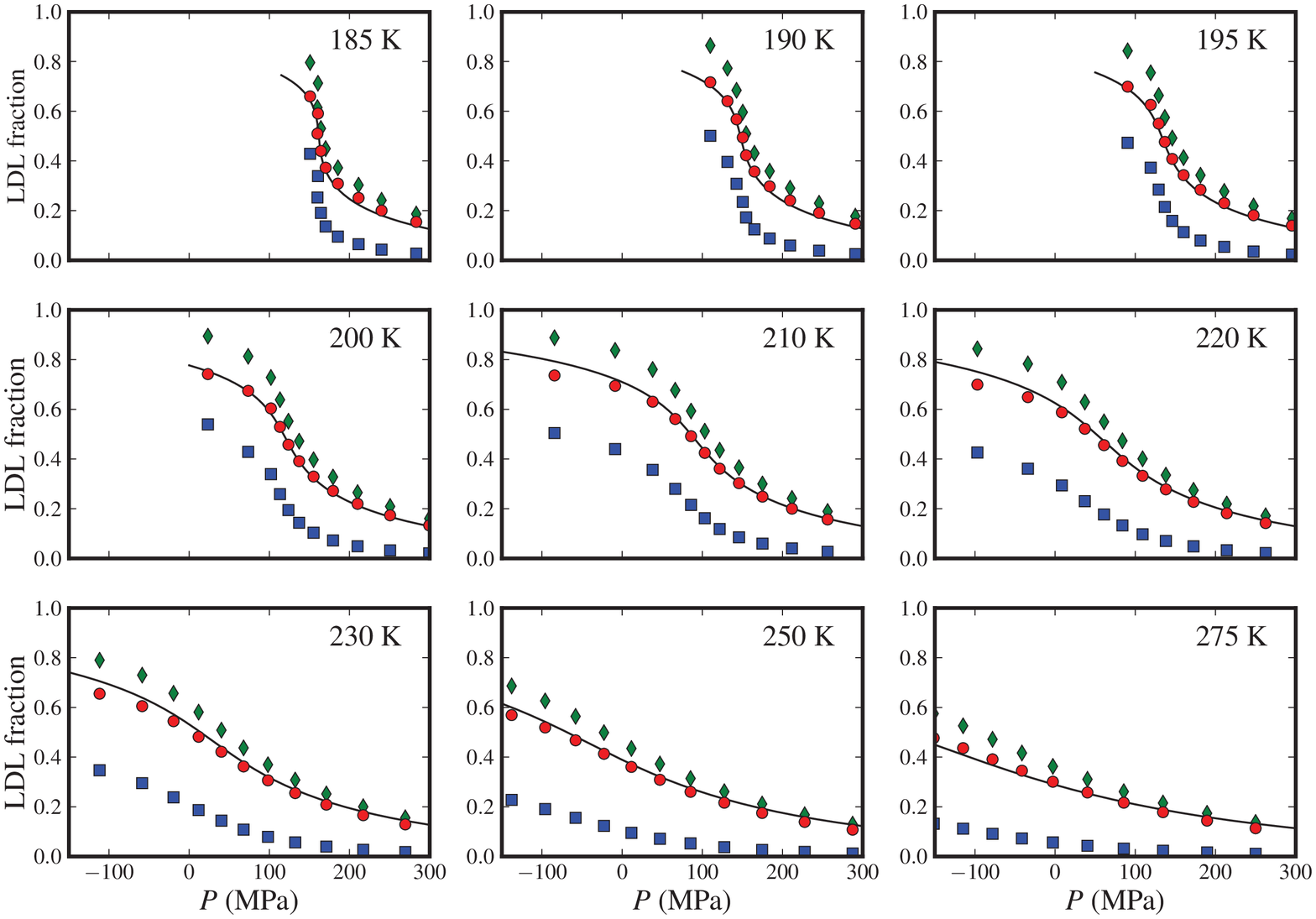}
\caption{Pressure dependence of low-density fraction computed using both $d_5$ (filled diamonds) and LSI (filled squares) criteria along different isotherms. Filled circles are $d_5$ multiplied by a factor of $0.82$. Solid lines indicate predictions of TSEOS for the low-density fraction.}
\label{fig5}
\end{figure}

In Fig.~\ref{fig5}, we compare the low-density fractions computed using both the $d_5$ and LSI criteria along with the predictions of the TSEOS for the extent of reaction, $x$. The computed order parameters and the phenomenological low-density fraction show qualitatively similar pressure dependence along different isotherms. The computed low-density fraction based on the LSI criterion is significantly lower than that based on the $d_5$ criterion.  LSI strongly underestimates the extent of reaction $x$, while $d_5$ slightly overestimates it. We also observe that the low-density fraction based on $d_5$ criteria is symmetric, showing an inflection point at about $1/2$, which is in agreement with the TSEOS definition of the order parameter that is related to the low-density fraction as $x-x_c$ with the critical fraction $x_c=1/2$. Low-density fractions computed for two versions of ST2~\cite{anisimov_st2} were also based on $d_5$ and were in good agreement with the TSEOS. Similar behavior is demonstrated by the local density structure order parameter introduced by Russo and Tanaka~\cite{tanaka_natcomm} which is also based on $d_5$-like criteria. 

Moreover, the low-density fraction, predicted by $d_5$ based criteria, multiplied by a factor $\sim 0.82$ is in remarkably good agreement with the phenomenologically defined order parameter.  The discrepancy between the LDL fraction obtained by the $d_5$ criterion and $(x-x_c)$ may originate from both the approximations made in the TSEOS and the details of the microscopic definition of the order parameter.  We also note that in the lowest approximation the phenomenological order parameter $x-x_c$ is proportional to the change in molar volume ($V$) and entropy ($S$) as $x-x_c=a\lambda(V-V_c)/V_c$ and as $x-x_c=-\lambda(S-S_c)/R$~\cite{anisimov_st2, anisimov_nat}.

\section{Phase Behavior OF TIP4P/2005 Water from TSEOS}\label{phase}
\begin{figure}[htbp!] 
\vspace{10pt}
\includegraphics[width=0.475\textwidth]{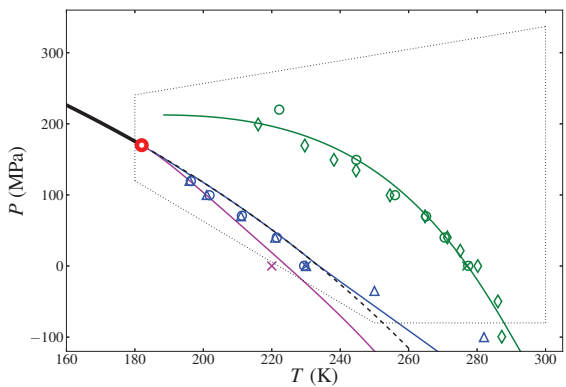}
\caption{Phase diagram for TIP4P/2005 water model predicted by TSEOS. The LLPT, critical point, and Widom line are shown by the thick black curve, the red circle, and the dashed black curve, respectively.  The loci of maxima in $\rho$, $\kappa_T$, and $C_P$ according to the TSEOS are shown by green, blue, and magenta curves, respectively.  Corresponding data are shown as reported by Refs. \cite{tanaka_natcomm} (open diamonds), \cite{abascal_2010} (open circles), \cite{caupin_pnas_2014} (open triangles), and as computed for this work (crosses). The dotted contour bounds the region of validity of the TSEOS.}
\label{fig6}
\end{figure}

In Fig.~\ref{fig6}, we present the phase diagram summarizing the behavior of supercooled TIP4P/2005 in the $P$-$T$ plane predicted by the TSEOS. In two-state thermodynamics, the locus of points with $\ln K = 0$ at $\omega > 2$ locates the LLPT line between HDL and LDL. The continuation of this line for $\omega < 2$ is the Widom line (see Section~\ref{tseos}). Using the language of the scaling theory of critical phenomena, $\ln K$ corresponds to the ordering field, while the conjugate variable $x-1/2$ is the order parameter.  The Widom line corresponds to zero field and zero order parameter and is the line of maximum fluctuations of the order parameter. Asymptotically, the Widom line coincides with the loci of the compressibility maxima and heat-capacity maxima (see also ref.~\cite{stanley_pnas_2005}). The estimated critical temperature, pressure and density are $T_c = 182$ K, $P_c = 170$ MPa, and $\rho_c = 1017$ kg/m$^3$, respectively. These critical parameters are in close agreement with the values reported by Sumi and Sekino~\cite{sumi_2013} and Yagasaki \textit{et al.}~\cite{yagasaki_2014}. The TMD line predicted from two-structure thermodynamics also
shows reasonable agreement with previously reported simulation data for this model~\cite{abascal_2010, tanaka_natcomm}. Our model predicts quite accurately (within $\sim 5$ K) the temperature of maximum of the isobaric heat capacity ($ \approx 220$ K~\cite{saito_bagchi}) and isothermal compressibility ($ \approx 230$ K~\cite{lars_2015}) at ambient pressure.              

In Fig.~\ref{fig6}, the dotted contour bounds the area of the validity of this form of the TSEOS. This restricted form becomes increasingly inaccurate for densities below $960$ kg/m$^3$ and at negative pressures. In the current work we did not consider pressures below $-80$~MPa. Extending the validity of the TSEOS to lower densities and negative pressures will at least require the restrictions on the definions of $\ln K$  and $\omega$ in Eqs.~\ref{eqn:3} and~\ref{eq5} to be relaxed.  Such an extension could address the current discussions surrounding the behavior of water at extremely strong negative pressures~\cite{caupin_pnas_2014}.

\section{Water-Like Models versus Real Water}

\begin{figure}[htbp!]
 \vspace{10pt}
        \subfigure{
            \label{fig7a}
            \includegraphics[width=0.42\textwidth]{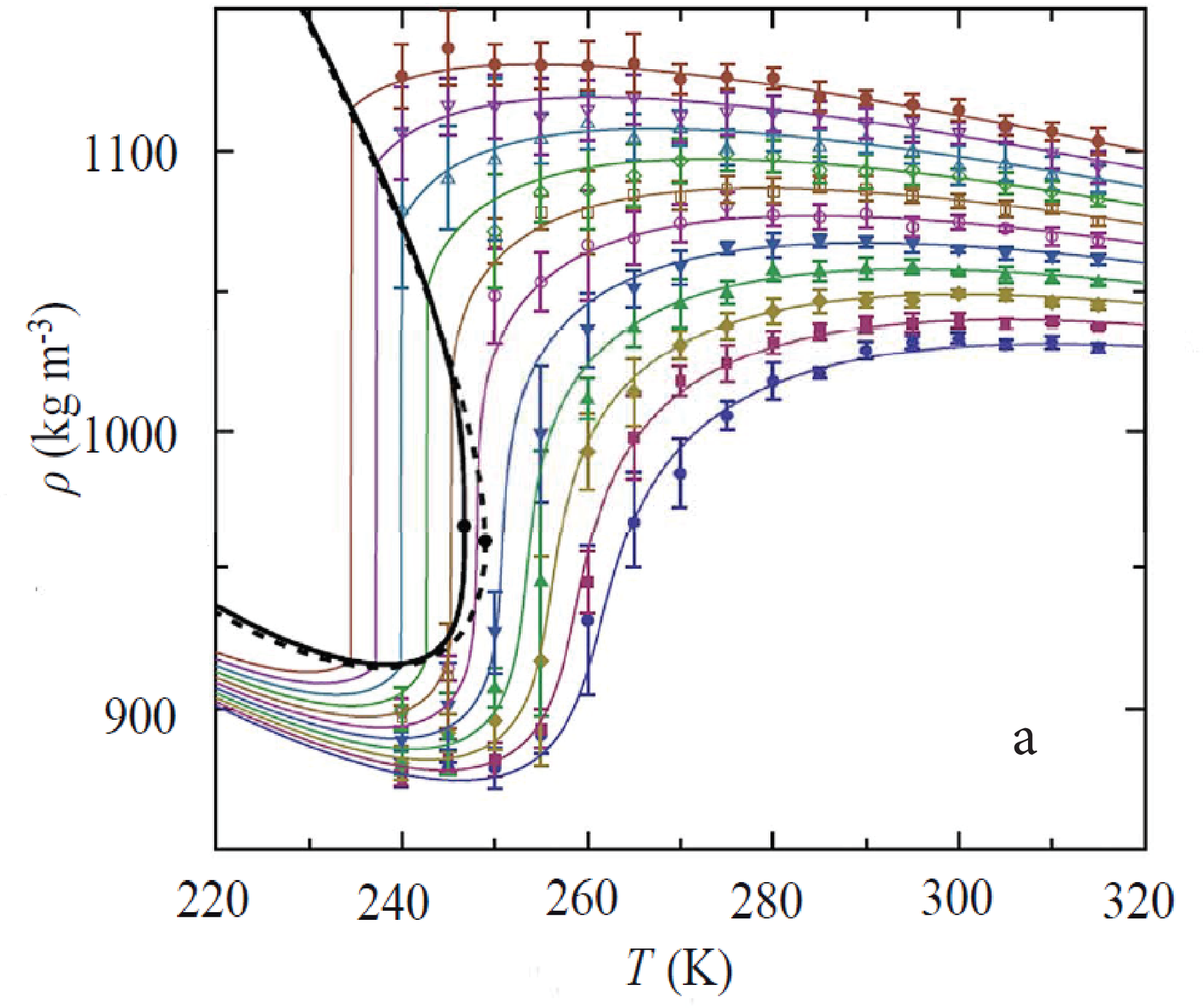}               
        }
        \subfigure{
           \label{fig7b}
           \includegraphics[width=0.47\textwidth]{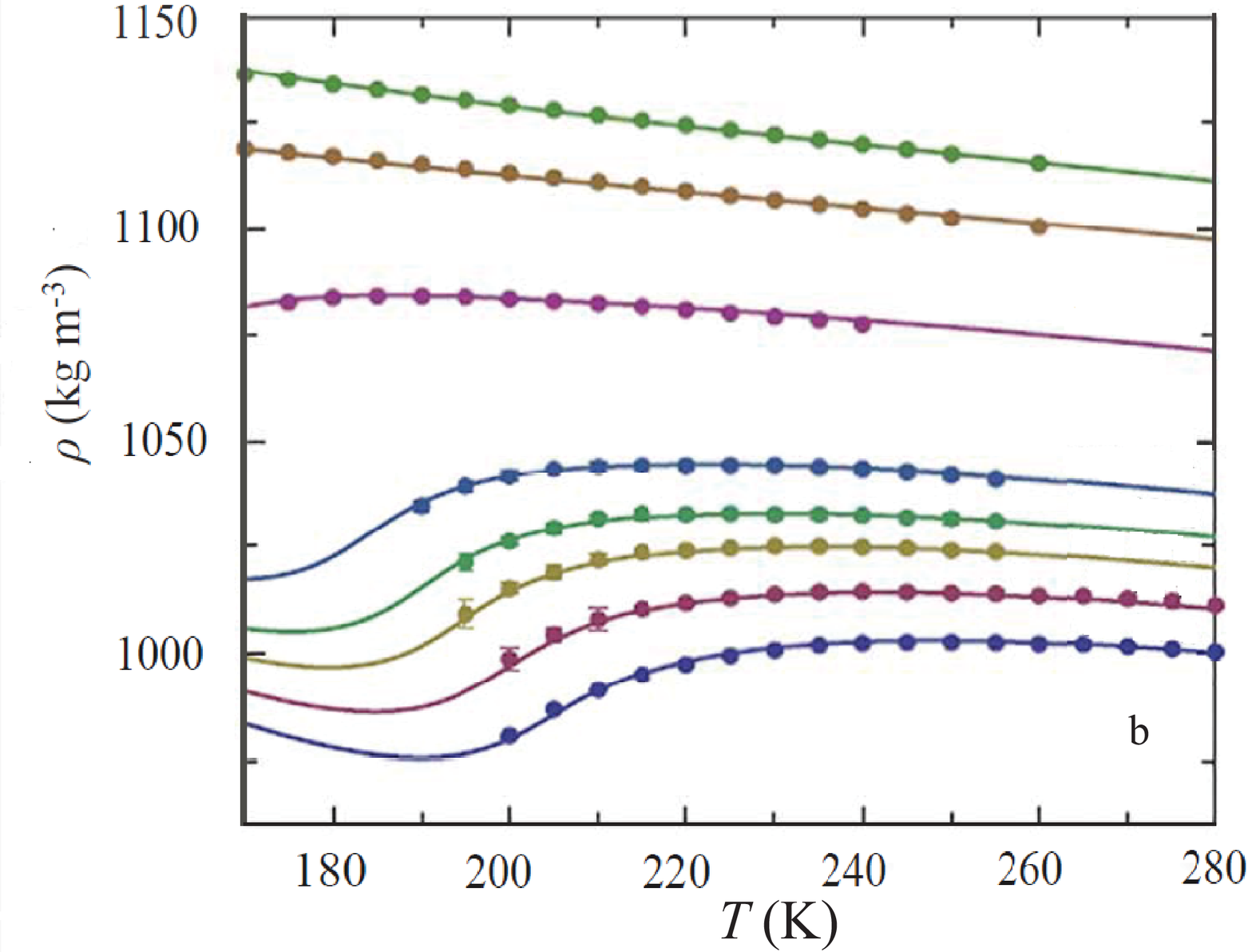}
        }
    \caption{Simulated data from the ST2 and mW model (symbols) are compared with the predictions (curves) of two-structure thermodynamics, as adapted for the respective models. (a) Temperature dependent density ($\rho$) along different isobars computed for (a) the ST2(II) model. The thick black curve indicates two-phase coexistence (dashed: mean field equation, solid: crossover equation) and black dots represent the critical point. The isobar pressures vary from $100$~MPa to $200$~MPa in steps of $10$ MPa. Figure adapted with permission from Ref.~\cite{anisimov_st2}, \textcircled{c} 2014, American Institute of Physics. (b) Temperature dependent density ($\rho$) along different isobars computed for the mW model.  Figure adapted with permission from Ref.~\cite{anisimov_mb}, \textcircled{c} 2013, American Institute of Physics.}
   \label{fig7}
\end{figure}

Our study, together with three previously published simulation results~\cite{abascal_2010, sumi_2013, yagasaki_2014}, shows that the TIP4P/2005 model in the range of pronounced thermodynamic anomalies behaves similarly to the ST2 model. This is clearly seen from the equally sharp behavior of isobars in the vicinity of the projected critical point as demonstrated in Figs.~\ref{fig7a} and ~\ref{fig2}. Even without computational data obtained for the two-phase region, such van der Waals-like behavior of the isobars suggests the proximity of the critical point. Contrary to the ST2 and TIP4P/2005 models, in the mW model of water the isobars, shown in Fig.~\ref{fig7b}, only weakly change with changing pressure and never become steep enough to suggest criticality. Indeed, the presence of a LLPT is model-dependent. While in the mW model the non-ideality in mixing of the two structures never becomes strong enough to cause a metastable LLPT, the thermodynamics of the ST2 and TIP4P/2005 models strongly implies the existence of a metastable LLPT.
\begin{figure}[htbp!]
 \vspace{10pt}
              \subfigure{
           \label{fig8a}
           \includegraphics[width=0.45\textwidth]{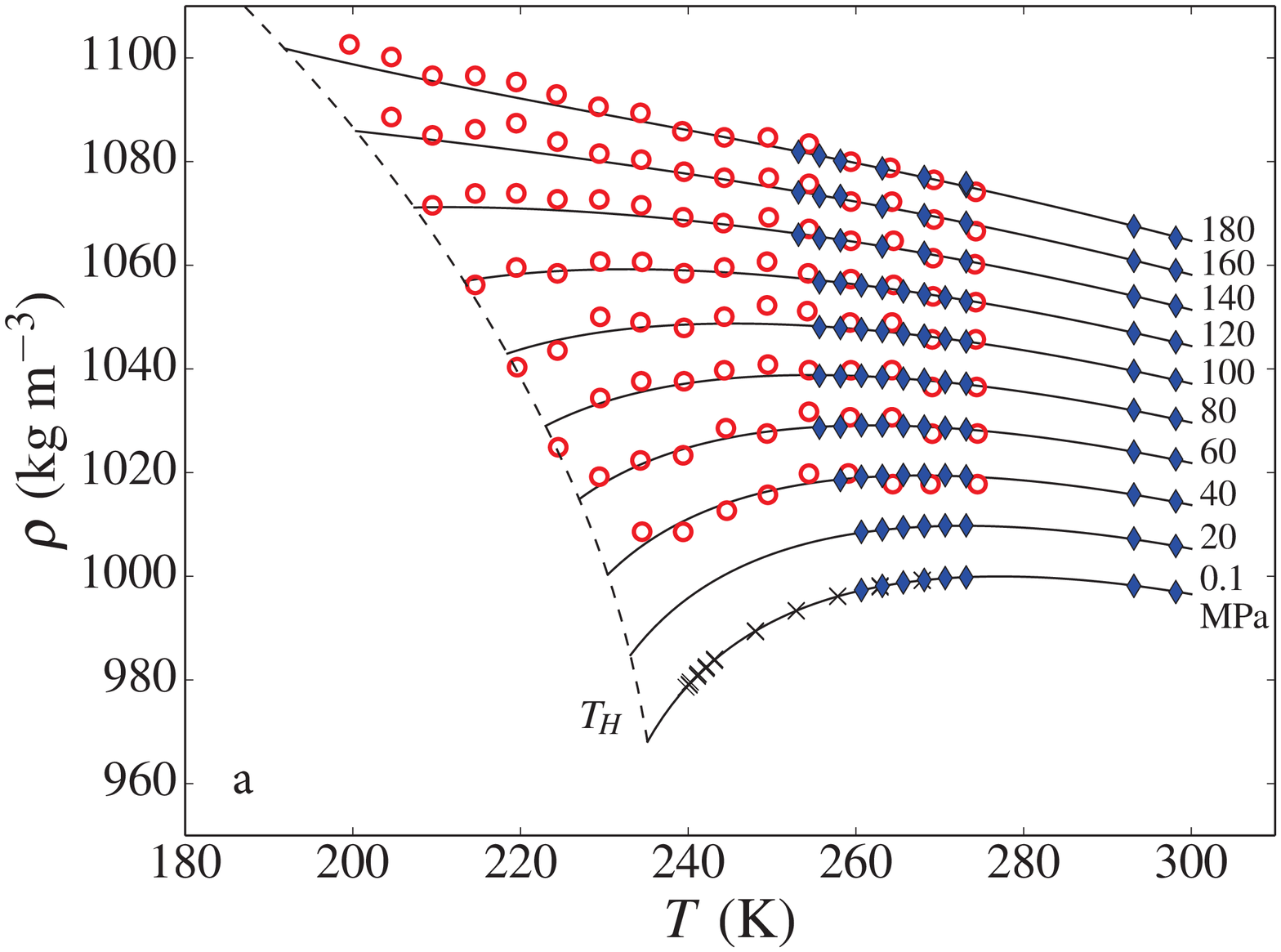}
        }  
            \subfigure{
           \label{fig8b}
           \includegraphics[width=0.45\textwidth]{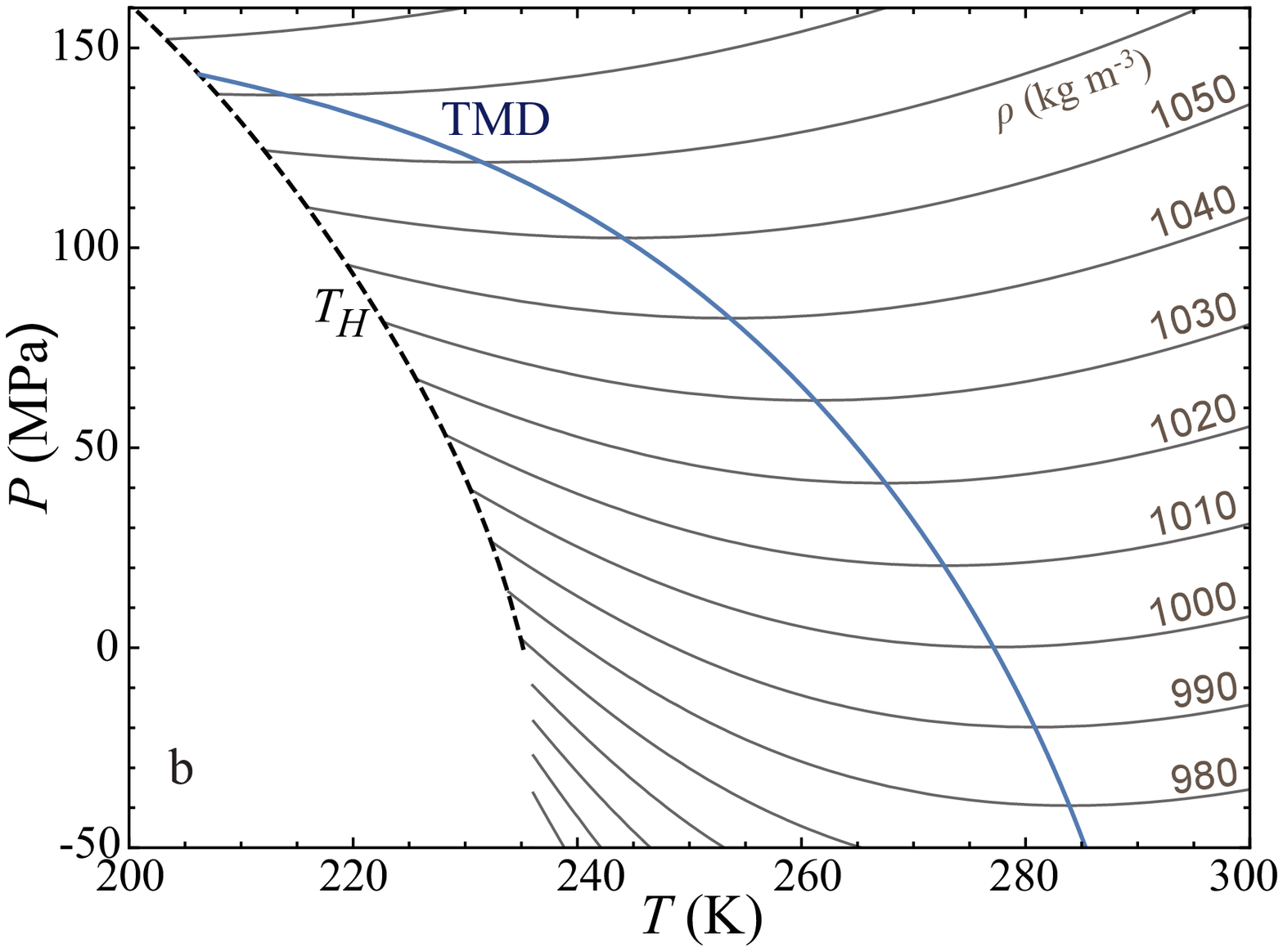}
        }
    \caption{(a) Density of cold and supercooled water as a function of temperature along different isobars (black lines are the predictions of an extended version of the TSEOS~\cite{anisimov_new}).  Symbols are experimental data reported in Refs.  \cite{hare_1987} (crosses), \cite{mishima_2010JCP} (open red circles), and \cite{sotani_2000} (filled blue diamonds). $T_H$ indicates the homogeneous nucleation line. The data from Ref. \cite{mishima_2010JCP} have been adjusted by at most 0.3\% to correct for small systematic errors, as explained in Ref. \cite{Holten_JCP2012}. (b) Isochores of cold and supercooled water computed with an extended version of the TSEOS~\cite{anisimov_new}. The dashed curve is the homogeneous nucleation line and the blue curve is the TMD locus.}
   \label{fig8}
\end{figure}

The thermodynamics of real supercooled water is more ambiguous. Properties of bulk supercooled water in the experimentally accessible region are well described by two-structure thermodynamics (for example, density data and theoretical predictions along isobars are presented in Fig.~\ref{fig8a}). However, the projected phase separation is located so far below the homogeneous ice nucleation limit that the location of a LLPT and even its very existence becomes uncertain~\cite{anisimov_nat}. This problem with real water is clearly illustrated by comparison of Fig.~\ref{fig1}, showing the convergence of isochores in TIP4P/2005 at a point that is interpreted as the LLCP, and Fig.~\ref{fig8b} for real water in which such convergence is in principle allowed but far from certain. Obviously, the real-water dilemma cannot be resolved with the experimental data that are currently available.  Future studies will need to either penetrate into \lq\lq no-man's land" or bring the critical point into experimentally accessible conditions by adding a solute~\cite{corradini_2010JCP,biddle_2014,zhao_angell_2016}.

\section{Discussion: Does a Metastable LLPT Exist in TIP4P/2005?}
We have investigated the thermodynamic behavior of the TIP4P/2005 water model in the supercooled region.  The convergence of the isochores around a density of about $1020$ kg/m$^3$ and the steep van der Waals-like behavior of the order parameter (the low-density fraction) at about $180-185$ K suggests the presence of a metastable LLPT in the TIP4P/2005 model.

Our results are supported by the data of Sumi and Sekino~\cite{sumi_2013} and consistent with the conclusions of Yagasaki \textit{et al.}~\cite{yagasaki_2014}.  The substantiation of this viewpoint will require free-energy calculations such as those that have yielded unambiguous evidence \cite{pablo_nature_2014,small_st2} of a liquid-liquid transition in the ST2 model of water.  Because the phenomenon under scrutiny is metastable, the question of how sampling time and system size constrain the possibility of observing a liquid-liquid transition arises in addition to the question of its existence in a free-energy or equation-of-state calculation.

However, the most recent extensive study of the TIP4P/2005 and TIP5P models by Overduin and Patey~\cite{patey_2015}, which reported simulations in the projected two-phase region for systems ranging in size from $4000$ to $32000$, found density differences between the regions of low and high densities to decrease with increasing system size. The difference finally disappeared for a system composed of $32000$ molecules.  Overduin and Patey further argued that, as the appearance of regions of low density is always accompanied by small ice-like crystallites, the regions of different densities observed by Yagasaki \textit{et al.}~\cite{yagasaki_2014} might be associated with the appearance and coarsening of local ice-like structures, rather than with liquid-liquid phase separation. This argument is similar to that of Limmer and Chandler~\cite{limmer_c}, who also argued that the density differences observed by Yagasaki \textit{et al.}~\cite{yagasaki_2014} are due to ice coarsening, rather than to spontaneous liquid-liquid phase separation. 

This argument deserves serious consideration. However, we must note that separated liquid states observed in $NVT$ simulations are always metastable with respect to ice formation.  Consequently, as Overduin and Patey note~\cite{patey_2015}, the mere presence of ice-like crystallites ($6-8\%$ for TIP4P/2005 model at the lowest temperature studied by Overduin and Patey~\cite{patey_2015}) having finite lifetime in the system does not provide unambiguous proof for the ice-coarsening hypothesis proposed by Limmer and Chandler~\cite{limmer_1, limmer_2, limmer_c}.  Also, the computed fraction of ice-like particles or crystallites is very sensitive to the definition adopted for classifying a water molecule as ice-like.  On the contrary, the observed excess local density of ice-like crystallites and strong correlations among them in low-density regions can also be understood without invoking the ice-coarsening hypothesis.  Liquid-liquid phase separation leads to spatial heterogeneity in water, and it is to be expected that the ice-like fluctuations or crystallites will be more stable in the low-density regions due to a lower surface free-energy cost.  In this context, the recent simulations of Smallenburg and Sciortino~\cite{small_st2} have clearly demonstrated that the liquid-liquid transition in the ST2 model is not a misinterpreted crystallization transition, as had been claimed~\cite{limmer_1, limmer_2}.

Moreover, the fact that phenomena observed in the metastable region depend on the system size and on the duration of observation time is not surprising.  This is, in fact, an essential characteristic of metastability. A metastable phase separation cannot exist at all in the thermodynamic limit (infinite size and infinite time).  If we denote by $\tau_{\mathrm{relax}}$ the internal reaxation time in the metastable state, and $\tau_{\mathrm{out}}$ the time it takes for the system to exit the metastable state and form the stable phase (\emph{i.\hspace{1mm} e.} a characteristic crystallization time in our case), then the metastable state is well defined if $\tau_{\mathrm{relax}} << \tau_{\mathrm{out}}$.  When this condition is met, thermodynamics can be applied to a metastable state. As Overduin and Patey note, there are several reasons why the metastable LLPT might not be manifested in large enough systems~\cite{patey_2015}.  In particular, as has been emphasized by Binder~\cite{binder_2014}, the divergence of the correlation length at the critical point causes the relaxation time to diverge, an effect known as critical slowing-down.  Increasing the system size, on the other hand, decreases the lifetime of metastability, and thus at certain conditions prevents the manifestation of metastable phase separation.  In addition, there is another characteristic timescale in this problem that could complicate observation of a liquid-liquid phase separation: the time of conversion between the two alternative liquid structures.  At temperatures well below the liquid-liquid  critical temperature (about 180~K) , this time may become long enough that the formation of the low-temperature structure will not be completed during the time of observation.

The formation of two liquid phases can also be impeded by the unfavorable interfacial energy between them.  Consequently, the extent of phase separation not only depends on the choice of initial density of the system but also on the aspect ratio of the simulation box. Due to the large surface energy cost for the formation of well-defined stable interfaces, phase separation is not observed in cubic boxes, even in systems far below the LLCP. In order to observe phase separation one always simulates rectangular boxes ($1:1:4$ in case of Yagasaki \textit{et al.}~\cite{yagasaki_2014} as well as Overduin and Patey~\cite{patey_2015} for $4000$ molecules) to minimize interfacial free energy cost for formation of the LDL-HDL interface.  It is thus very plausible that the observation of two different metastable liquid densities in water-like models, such as TIP4P/2005 and TIP5P, would involve length and time scale constraints that would also influence the pathway to homogeneous ice nucleation.

As explained above, attempts to directly observe metastable liquid-liquid separation in $NVT$ simulations are subject to non-trivial limitations.  We have used an alternative approach to evaluate the hypothesis of the metastable LLPT in supercooled water. We have studied a relatively small system of hundreds of molecules and performed a series of simulations (about $200$) to obtain reliable information on the thermodynamic surface. Our study does not support one of the scenarios discussed by Overduin and Patey~\cite{patey_2015} in which ``liquid-liquid coexistence is simply not a possibility" for the TIP4P/2005 water model. On the contrary, the clear convergence of the isochores around $1020$ kg/m$^3$ and the behavior of thermodynamic properties demonstrate the tendency to criticality. Furthermore, the equation of state that is built on the assumption of the existence of LLPT fits the simulation data very well. Moreover, the microscopic structural order parameters ($d_5$ and LSI) associated with the low-density fraction, strongly support the two-structure nature of TIP4P/2005 and the approach to criticality around $180$-$182$ K.  This important simulation result is independent of any speculation regarding the shape of the thermodynamic surface.

An alternative hypothesis to the competition between two liquid structures would be to attribute supercooled water anomalies (the sharp increases of the response functions) to pre-crystallization effects~\cite{limmer_c, patey_2015}. Indeed, the theory of so-called \lq\lq weak crystallization", which accounts for translational-order fluctuations, describes the properties of the supercooled mW model as well as two-structure thermodynamics does~\cite{anisimov_mb}. However, pre-crystallization effects cannot explain the convergence of isochores and the critical-like behavior of the low-density fraction that is clearly observed in the ST2 and TIP4P/2005 models.

There is another puzzling result of Overduin and Patey~\cite{patey_2015} that requires further studies. The correlation length characterizing fluctuations of density increases sharply upon supercooling in real water~\cite{Huang_2010, Wikfeldt_2011b}.  In Ref.~\cite{patey_2015}, Overduin and Patey examine this correlation length in both TIP5P and TIP4P/2005 and claim that it apparently diverges along the critical isochore in TIP5P, but does not exhibit such an anomaly in TIP4P/2005. We note that in our simulations the isothermal compressibility increases by an order of magnitude along the critical isochore, which is a strong effect, especially in view of a relatively small size of the system (about $2$ nm). The correlation length of density fluctuations is approximately proportional to the square root of the compressibility. Accordingly, the correlation length should increase by about three times, the effect indeed observed for TIP5P~\cite{patey_2015}.

In conclusion, the results of our study strongly support the presence of a liquid-liquid critical point in the TIP4P/2005 model, and are consistent with the possiblity of a liquid-liquid phase transition for this model. Our study does not answer the questions regarding conditions under which the metastable LLPT can or cannot be observed in the region below the projected critical point. Systematic studies at various simulation conditions are required to further our understanding of this deep and important problem. As far as the one-phase metastable liquid region is concerned, investigation of finite-size effects on the shape of the thermodynamic anomalies would be highly desirable.

\begin{acknowledgments}
JWB and MAA thank Frederic Caupin for fruitful discussions and his hospitality during part of this project, Valeria Molinero for useful comments, and Vincent Holten for valuable consultation. PGD gratefully acknowledges the support of the National Science Foundation (Grants No. CHE-1213343 and CBET-1263565).
\end{acknowledgments}


\bibliography{tip4p}

\onecolumngrid
\appendix
\section{Behavior of the self-intermediate scattering function in the deeply supercooled region}

Atomistic models of water are well known for extremely slow structural relaxation in the deeply supercooled state, especially in the low-temperature and low-density region of the phase diagram. To establish conclusively that our molecular dynamics (MD) trajectories are long enough to ensure the structural relaxation of the system even in the deeply supercooled region, we show in Fig. S1 the decay of the self-intermediate scattering function $F_s(k^*,t)$ ($k^*$ is the wavenumber corresponding to the first peak of structure factor) with time in the low-temperature and lower-density (near and below the liquid-liquid critical density, $\rho_c$ = 1017~kg/m$^3$ ) region of the phase diagram. Fig. S1(a) describes the decay of the self-intermediate scattering function at several temperatures on a near-critical isochore ($\rho$ = 1020 kg/m$^3$), and Fig. S1(b) describes the same on the $\rho$ = 980~kg/m$^3$ isochore (the lowest density at which we could relax the system up to the close vicinity of the critical temperature, $T_c$ = 182~K). It is quite evident from the figure that, even in the deeply supercooled region, the structural relaxation times and MD simulation run lengths (10 microseconds in both cases) are well separated. We also did not observe any sign of crystallization during our MD simulation at the reported thermodynamic conditions.  

\begin{figure}[h!]
 \vspace{5pt}
       \subfigure{
           \label{S1a}
           \includegraphics[width=0.45\textwidth]{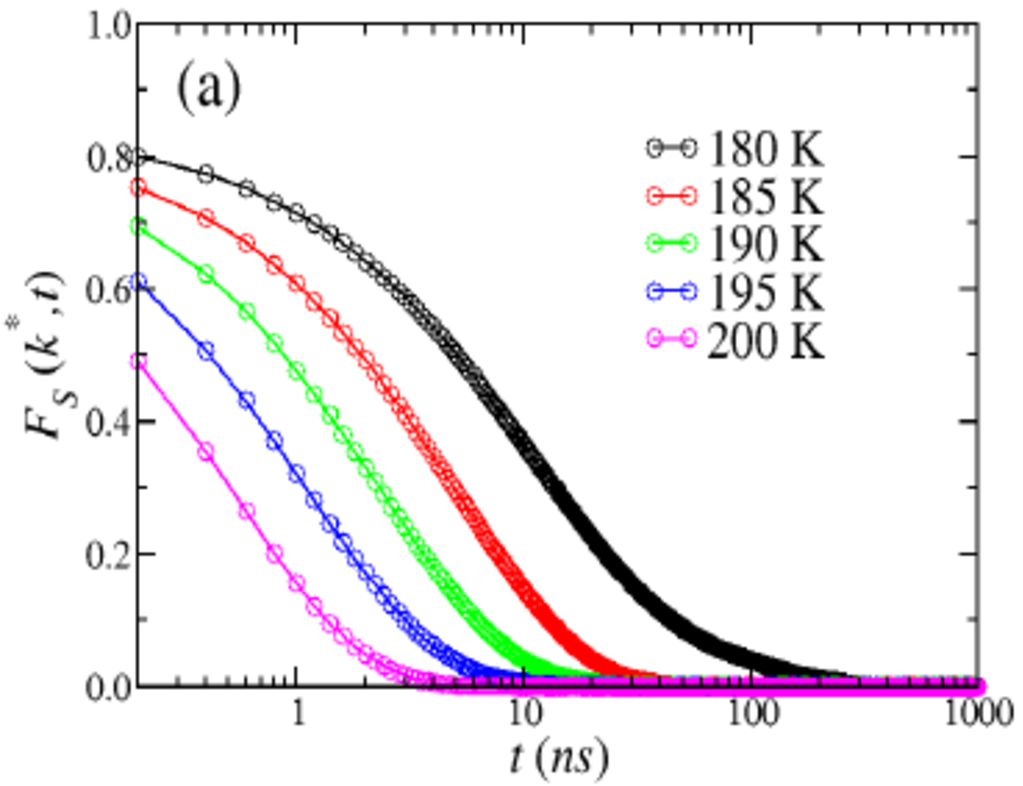}
        }  
           \subfigure{
           \label{S1b}
           \includegraphics[width=0.45\textwidth]{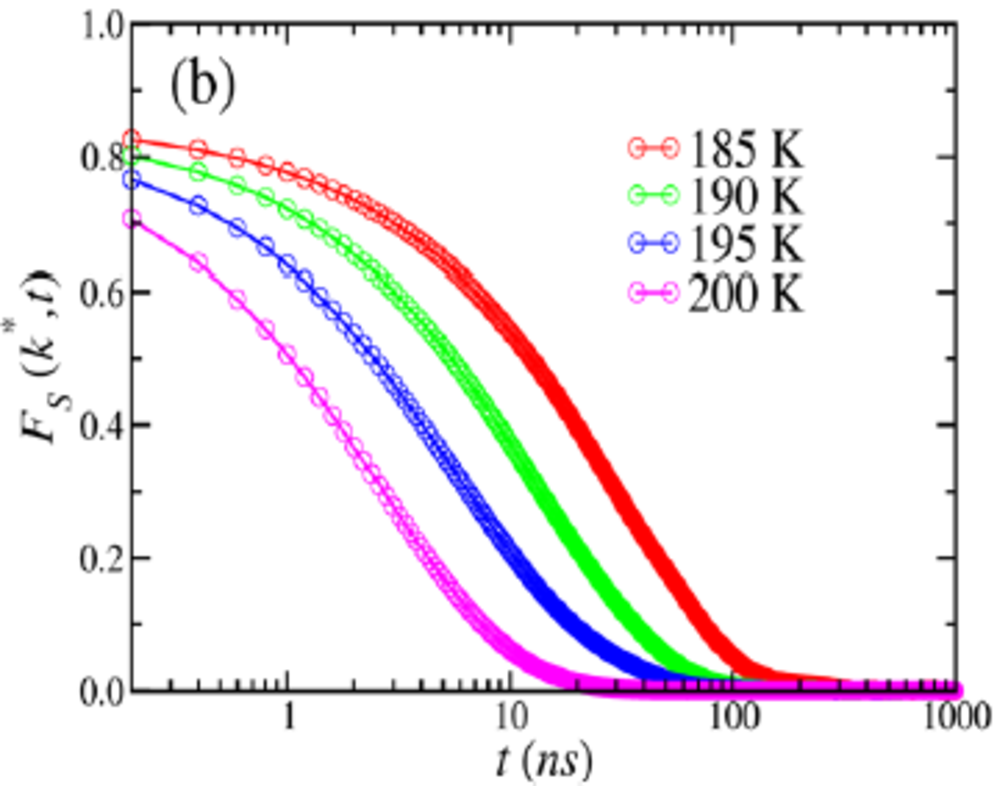}
        }
    \caption{The decay of the self-intermediate scattering function ($F_s(k^*,t)$, where $k^*$ is the wavenumber corresponding to the first peak of structure factor) with time at different temperatures in the deeply supercooled region along isochores: (a) 1020 kg/m$^3$  (close vicinity of the critical isochore, $\rho_c$ = 1017 kg/m$^3$ ) and (b) 980 kg/m$^3$  (the lowest density at which we could relax our system up to the close vicinity of the critical temperature, $T_c$ = 182~K). The MD simulation trajectory lengths at these conditions are 10 microseconds.}
\end{figure}

\end{document}